\documentclass[conference,10pt]{IEEEtran}

\usepackage[cmex10]{amsmath}
\usepackage{psfrag}

\ifCLASSINFOpdf

\else

\fi

\usepackage{graphicx}
\usepackage{subcaption}

\makeatletter
\def\endthebibliography{%
	\def\@noitemerr{\@latex@warning{Empty `thebibliography' environment}}
	\endlist
}

\makeatother

\begin{document}

\title{Channel Hardening in Massive MIMO ---\\A Measurement Based Analysis}

\author{\IEEEauthorblockN{Sara Gunnarsson\IEEEauthorrefmark{1}\IEEEauthorrefmark{2},
		Jose Flordelis\IEEEauthorrefmark{1},
		Liesbet Van der Perre\IEEEauthorrefmark{1}\IEEEauthorrefmark{2} and
		Fredrik Tufvesson\IEEEauthorrefmark{1}}
	\IEEEauthorblockA{\IEEEauthorrefmark{1}Department of Elecrical and Information Technology, Lund University, Sweden}
	\IEEEauthorblockA{\IEEEauthorrefmark{2}Department of Electrical Engineering, KU Leuven, Belgium}
	\IEEEauthorblockA{Email: \{sara.gunnarsson, jose.flordelis, fredrik.tufvesson\}@eit.lth.se, liesbet.vanderperre@kuleuven.be}}

\maketitle

\begin{abstract}
Wireless-controlled robots, cars and other critical applications are in need of technologies that offer high reliability and low latency. Massive MIMO, Multiple-Input Multiple-Output, is a key technology for the upcoming 5G systems and is one part of the solution to increase the reliability of wireless systems. More specifically, when increasing the number of base station antennas in a massive MIMO systems the channel variations decrease and the so-called channel hardening effect appears. This means that the variations of the channel gain in time and frequency decrease.
In this paper, channel hardening in massive MIMO systems is assessed based on analysis of measurement data. For an indoor scenario, the channels are measured with a 128-port cylindrical array for nine single-antenna users. The analysis shows that in a real scenario a channel hardening of $\mathbf{3.2}$--$\mathbf{4.6}$~dB, measured as a reduction of the standard deviation of the channel gain, can be expected depending on the amount of user interaction. Also, some practical implications and insights are presented.
\end{abstract}


\IEEEpeerreviewmaketitle

\section{Introduction}
One goal for future wireless communication systems is to support critical communications, meaning that low latency and high reliability are required. One of the key technologies to reach this goal is massive MIMO, massive Multiple-Input Multiple-Output \cite{6736761}. By increasing the number of antennas at the base stations and exploiting the spatial diversity, massive MIMO systems can achieve higher reliability. One reason for this is the channel hardening effect, which becomes present when many antennas are deployed. With channel hardening, fast-fading decreases and the channel starts to behave almost deterministically. This means that the problem with small-scale fading decreases and leaves only the large-scale fading to handle, which simplifies the channel estimation and power allocation among other things. 

There are two channel hardening effects associated with massive MIMO. Firstly, the experienced delay spread is decreased, which means that the fading over frequency becomes small or even negligible. Secondly, the fading in time decreases due to the coherent combining of the signals from the many base station antennas. Channel hardening was theoretically dealt with in \cite{Marzetta2010}\cite{Bjoernson2017}\cite{7880691}. The temporal fading was analyzed using massive MIMO measurements in \cite{7869694}, whereas the root-mean-square delay spread as a function of the number of antennas in a measured massive MIMO system was investigated in \cite{6666103}. 

In this paper, channel hardening in massive MIMO is analyzed based on data from a measurement campaign taking place in an indoor auditorium with a 128-port cylindrical array and nine closely-spaced users. The target of our analysis is to gain insights about how channel hardening is experienced in a practical scenario.

The structure of the paper is as follows. Section~II describes the measurement scenario and the measurement equipment. Thereafter, in Section~III, the theory which the analysis is based on is given. Provided the necessary background, Section~IV presents the results where the data from the measurements is analyzed with the previously given theory as a foundation. Finally, some conclusions are presented. 

\section{Measurement scenario and equipment}
\begin{figure}[t]
	\centering
	\includegraphics[width=3in]{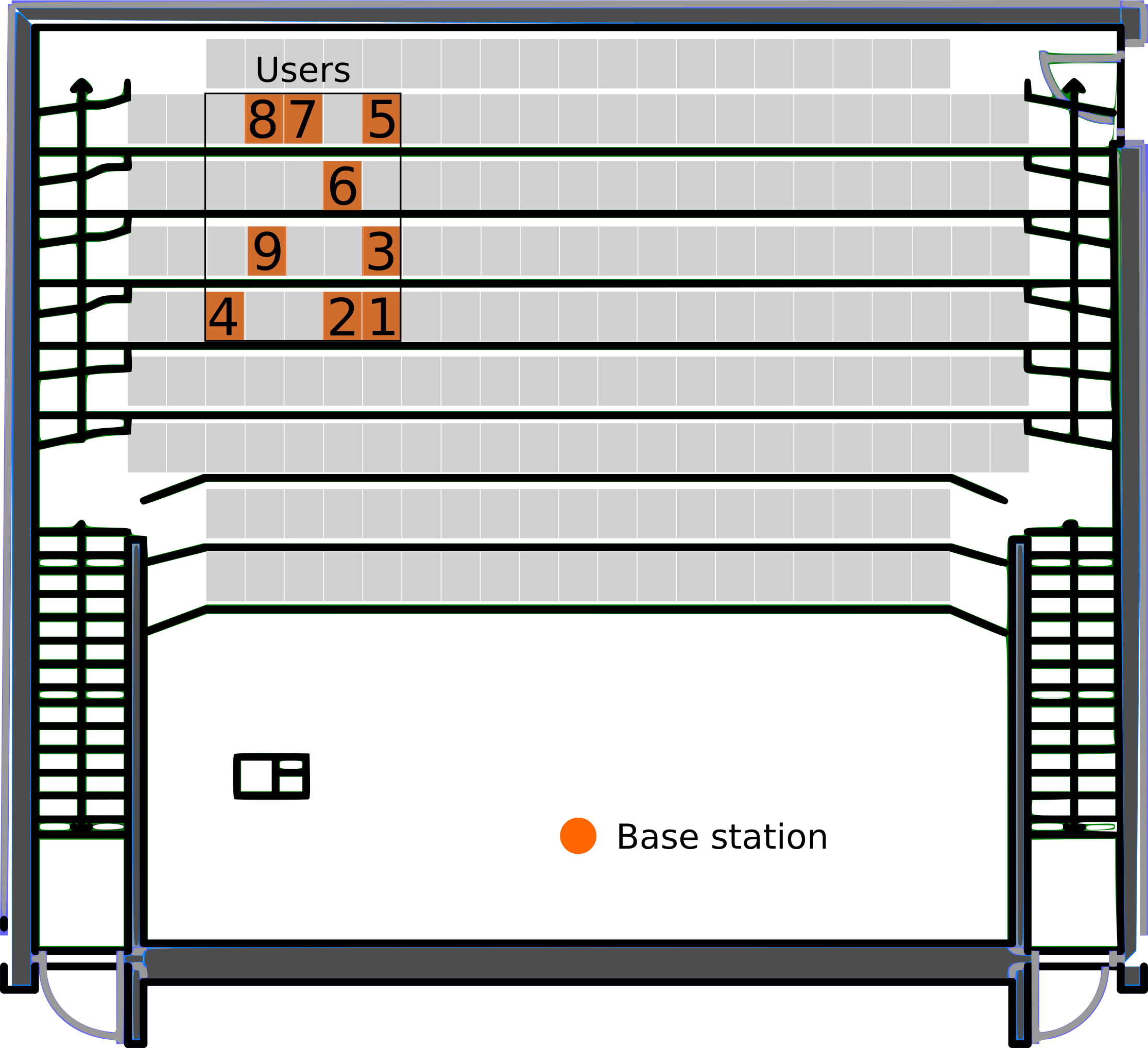}
	\caption{Floor plan of the room where the measurements took place. The base station is standing in the front of the room and the users are seated in the back of the room to the left.}
	\label{fig:location}
\end{figure}

The analysis is based upon data coming from a measurement campaign, detailed in \cite{Bourdoux2015}. The scenario considered is an indoor auditorium at Lund University with one base station and nine closely-spaced users placed as described in Fig.~\ref{fig:location}. The room is about $15.0\times12.4\times5.8$~meters. The users are placed at seats spread over four rows and five columns in the auditorium and are mostly static but with some slow movements, up to 1~m/s, and hold the antennas tilted 45~degrees. Line-of-sight (LOS) propagation conditions predominate, with occasional blocking due to other users or room furniture.

\begin{figure}[t]
	\centering
	\includegraphics[width=3.5in]{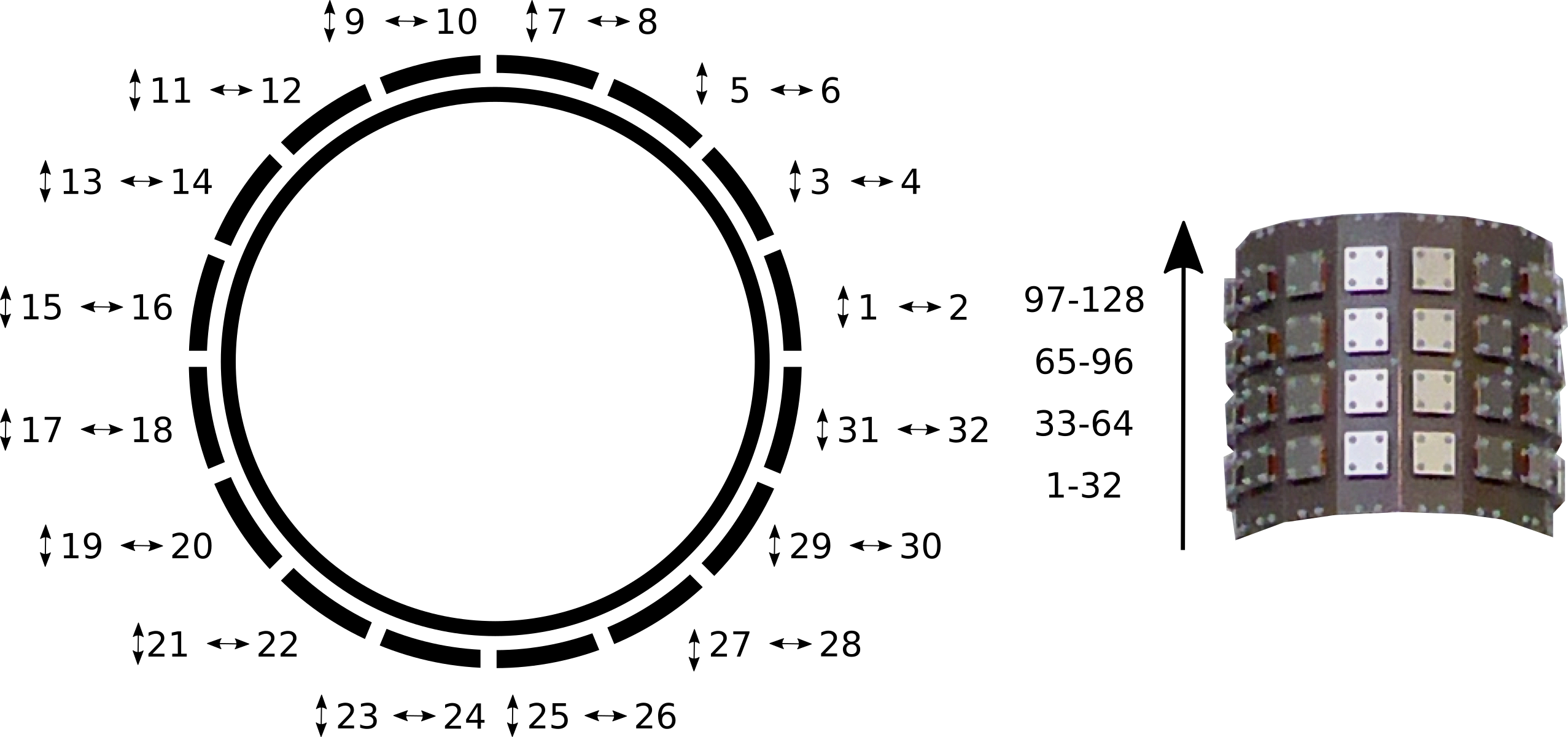}
	\caption{The base station antenna array as seen from above (left) with the numbering of the antenna elements in the first ring, both vertically and horizontally polarized. The cylindrical array seen from the side (right) with the numbering per ring.}
	\label{fig:array}
\end{figure}

The RUSK LUND MIMO channel sounder used in the measurements is a multiplexed-array channel sounder, which transmits OFDM-like symbols and measures the transfer function for all transmit-receive antenna combinations rapidly after each other. The receive unit of the channel sounder acts as a base station, and is equipped with a 128-port cylindrical array consisting of 64 dual-polarized patch antennas, spaced half a wavelength apart. The antenna elements are distributed in four rings on top of each other, as shown in Fig.~\ref{fig:array}. The array is situated at the height of 3.2~meters and 1.85~meters from the wall. The antennas utilized by the nine users are of the type SkyCross SMT-2TO6MB-A, which are omni-directional antennas with vertical polarization.
The measurements were taken when the base station was communicating with the nine users at a center frequency of 2.6~GHz and a bandwidth of 40~MHz, resulting in 129~measured points in frequency and 300~snapshots taken over 17~seconds. The antenna elements are in the measurement data numbered according to Fig.~\ref{fig:array}, i.e. starting from the lowest of the four rings and then going upwards. For each ring, the numbering starts with the antennas pointing to the right in Fig.~\ref{fig:location}, where the antennas with vertical polarization have odd numbers. Then, the numbering continues counter-clockwise in the ring before moving up to the next ring, and so on.

\section{Channel hardening}
The definition of channel hardening used follows \cite{7880691}, where it is considered that a channel $\mathbf{h}_{k}$ offers hardening if

\begin{equation}
\frac{\text{Var}\{\|\mathbf{h}_{k}\|^2\}}{(\text{E}\{\|\mathbf{h}_{k}\|^2\})^2}\rightarrow 0, \hspace{0.5cm} \text{as} \hspace{0.2cm} M\rightarrow \infty,
\end{equation}

\noindent where $\mathbf{h}_k$ is the channel vector for user $k$ and $M$ is the number of base station antennas. In this paper, the standard deviation, i.e. the square-root of the variance, is used as the metric of interest. Therefore, the investigation here concerns the standard deviation of channel gain for different numbers of base station antennas, similar to \cite{7869694}.

Starting with the normalization, the measured channel transfer functions have been normalized according to

\begin{equation}
\mathbf{\overline{h}}_{k}(n,f) = \frac{\mathbf{h}_{k}(n,f)}{\sqrt[]{\frac{1}{N F M}\sum_{n=1}^{N}\sum_{f=1}^F\sum_{m=1}^M|h_{km}(n,f)|^2}},
\label{eq:norm}
\end{equation}



\noindent where $N$ is the number of snapshots, $F$ is the number of frequency points and $M$ is the number of \textit{selected} base station antennas. This normalization makes sure that the average power of each entry in $\overline{\mathbf{h}}_{k}$, averaged over frequency, time, and base station antennas, is equal to one.

For $M$ \textit{selected} base station antennas, the instantaneous channel gain for each user is defined as

\begin{equation}
\overline{G}_k(n,f) = \frac{1}{M}\sum_{m=1}^{M} |\overline{h}_{km}(n,f)|^2,
\label{eq:subset_pwr}
\end{equation}

\noindent meaning that the mean channel gain

\begin{equation}
\mu_k = \frac{1}{N F} \sum_{n=1}^N \sum_{f=1}^F \overline{G}_k(n,f),
\label{eq:mean}
\end{equation}

\noindent is independent of the number of antennas selected at the base station. The total output power of the base station can hence be reduced with a factor corresponding to the beamforming gain, $M$. Note that this normalization is slightly different from the one used in \cite{7869694}. Finally, the standard deviation of channel gain is computed for each user according to

\begin{equation}
Std_k = \sqrt{\frac{1}{N F} \sum_{n=1}^{N} \sum_{f=1}^{F} |\overline{G}_k(n,f) - \mu_k|^2},
\label{eq:std}
\end{equation}

\noindent where the mean channel gain, $\mu_k$, for user $k$ is given in (\ref{eq:mean}).

\section{Analysis}
The data from the measurement campaign was analyzed in order to pinpoint the properties that create channel hardening. The following analysis visualizes and discusses some characteristics which can be found in an actual MIMO channel.

\begin{figure}[t]
	\centering
	\includegraphics[width=3.75in]{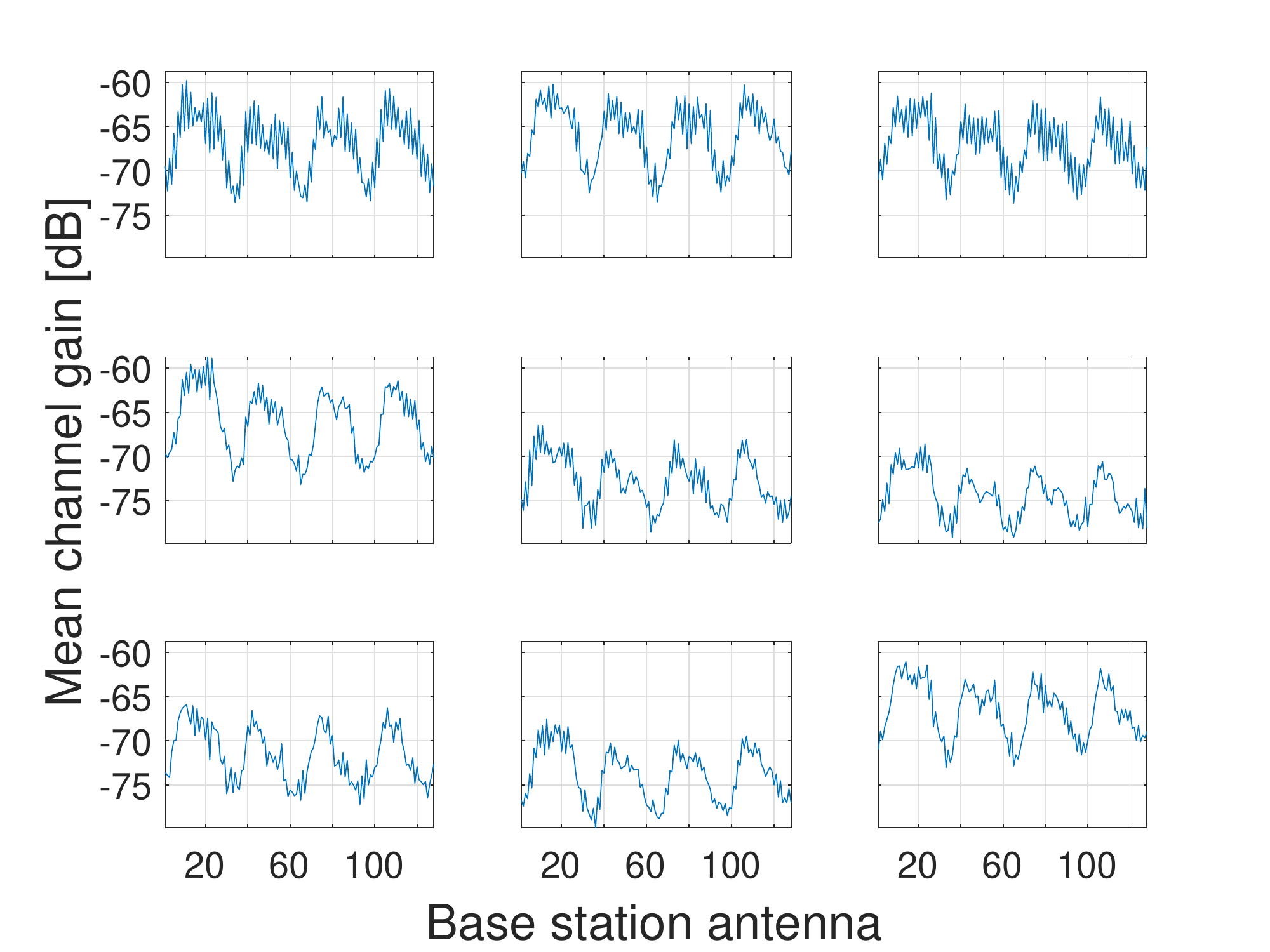}
	\caption{Mean channel gain over the base station array with 128~base station antenna shown for each of the nine users. The channel is not normalized. Users are numbered starting at the first row and column and then row-wise from left to right.}
	\label{fig:mean_pwr}
\end{figure}

In Fig.~\ref{fig:mean_pwr}, the mean unnormalized channel gain over the base station array with 128~base station antennas is shown. The reason for starting with the unnormalized channel gain is to show what the gain really can look like in practice, with the differences among antennas and the differences between users. The gain is shown for all nine users, starting with the plot in row one, column one and then going row-wise from left to right. What can be seen is that there are large differences in the mean channel gain, i.e. the gains for the channel scalars $h_{km}$ averaged over the array, for the different users. The users seated on the two lower rows (see Fig.~\ref{fig:location}), i.e. user~1, 2, 3, 4 and 9, have a higher mean channel gain in relation to the other users. As a comparison, the largest difference of mean channel gain between two users is between user~4 and user~6, where the difference is $8.9$~dB.

\begin{figure}[t]
	\centering
	\includegraphics[width=3.75in]{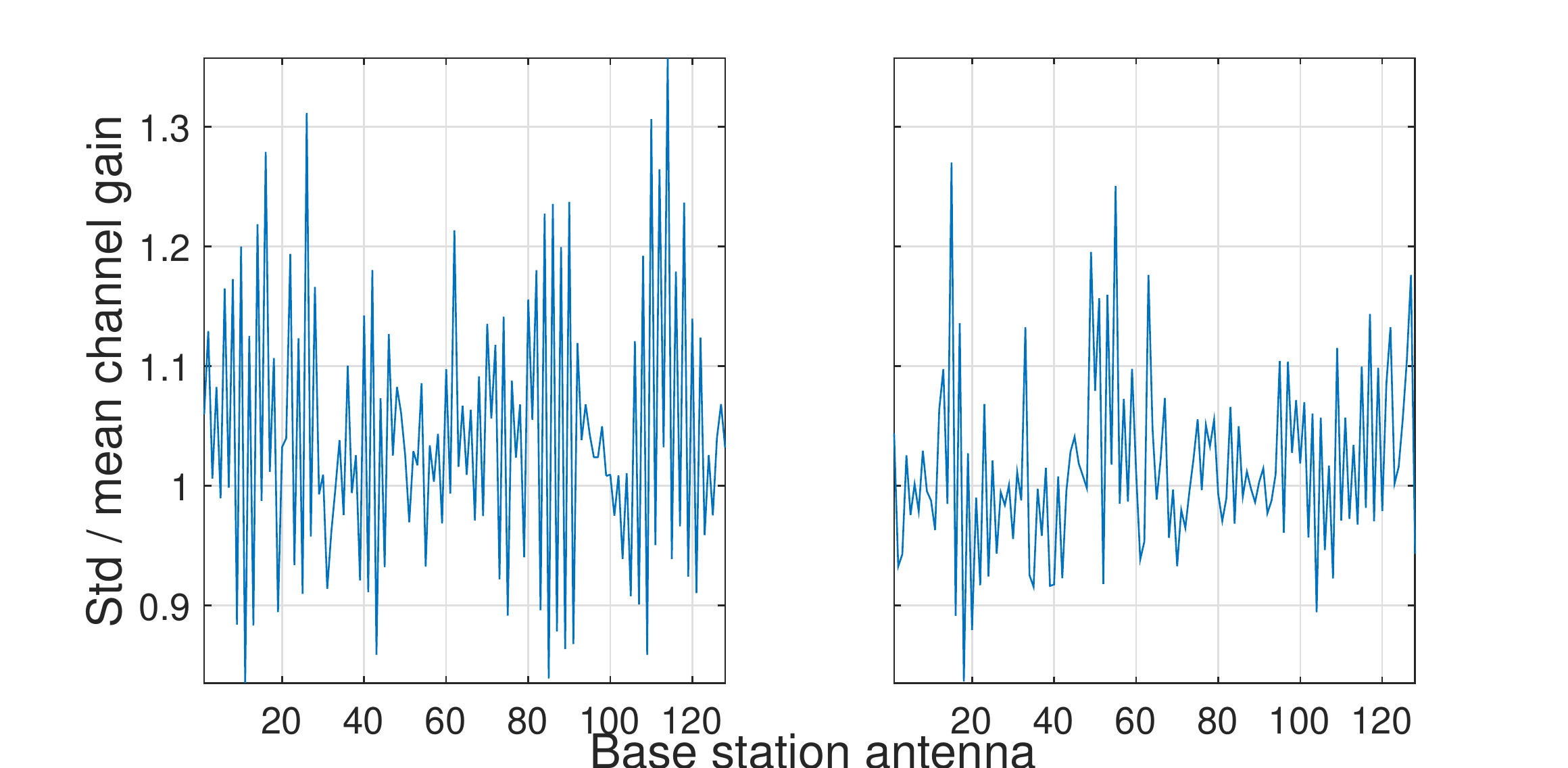}
	\caption{Standard deviation of channel gain divided by mean channel gain for user~1 (left) and 5 (right), i.e. the user in the uppermost left corner and in the middle of Fig.~\ref{fig:mean_pwr}.}
	\label{fig:std_pwr}
\end{figure}

What also can be seen in Fig.~\ref{fig:mean_pwr} is that variations over the base station array for each user also are very noticeable, where the difference between the peak value and bottom value for one user is varying between $10.6$~dB and $14.4$~dB. One reason causing these large variations over the array is the fact that it is a cylindrical array, meaning that some antenna elements experience a LOS condition while some do not. The particular antenna element's position, numbered according to Fig.~\ref{fig:array} and described in Section~II, explains the alternating behavior with the four peaks and dips. There is also an alternating behavior locally between two consecutive antennas. This variation is due to the fact that every other antenna element is vertically polarized and the other ones have a horizontal polarization. Also, the actual behavior of an antenna element is determined by its antenna pattern.

For the remaining analysis two representative users, user~1 and user~5, are selected. The reason for this choice is that user~1 is placed in front of the group and user~5 in the back and user~1 has larger differences in channel gain between the two polarizations while it is more equal for user~5. The standard deviation of channel gain for each base station antenna is computed according to (\ref{eq:std}). The standard deviation is then normalized with the mean channel gain for the targeted user to be able to better compare the standard deviation of channel gain for the two users since they have different means. In Fig.~\ref{fig:std_pwr} the results of this can be seen, user~1 being to the left and user~5 to the right. Especially for user~1, it can be seen that the largest standard deviation is when the antenna is in LOS where it alternates between the two polarizations.

\begin{figure}[t]
	\psfrag{Snap}[][]{\scriptsize{Snapshot}}
    \psfrag{Freq}[][]{\scriptsize{Frequency}}
	\centering
	\includegraphics[width=3in]{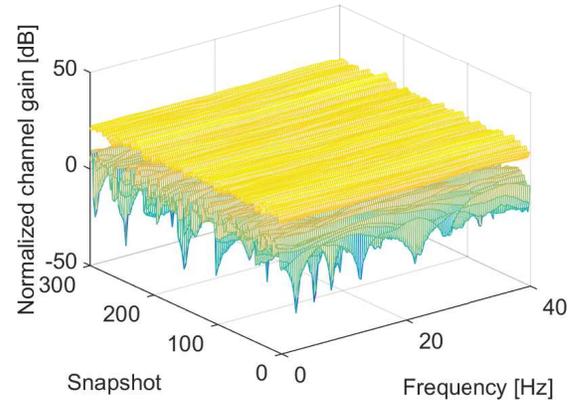}
	\caption{Normalized channel gain for one (lower) or 128~base station antennas (upper), respectively. The single base station antenna used is the one with the highest mean channel gain for user~1.}
	\label{fig:3D_best_1}
\end{figure}

\begin{figure}[t]
	\centering
	\includegraphics[width=3in]{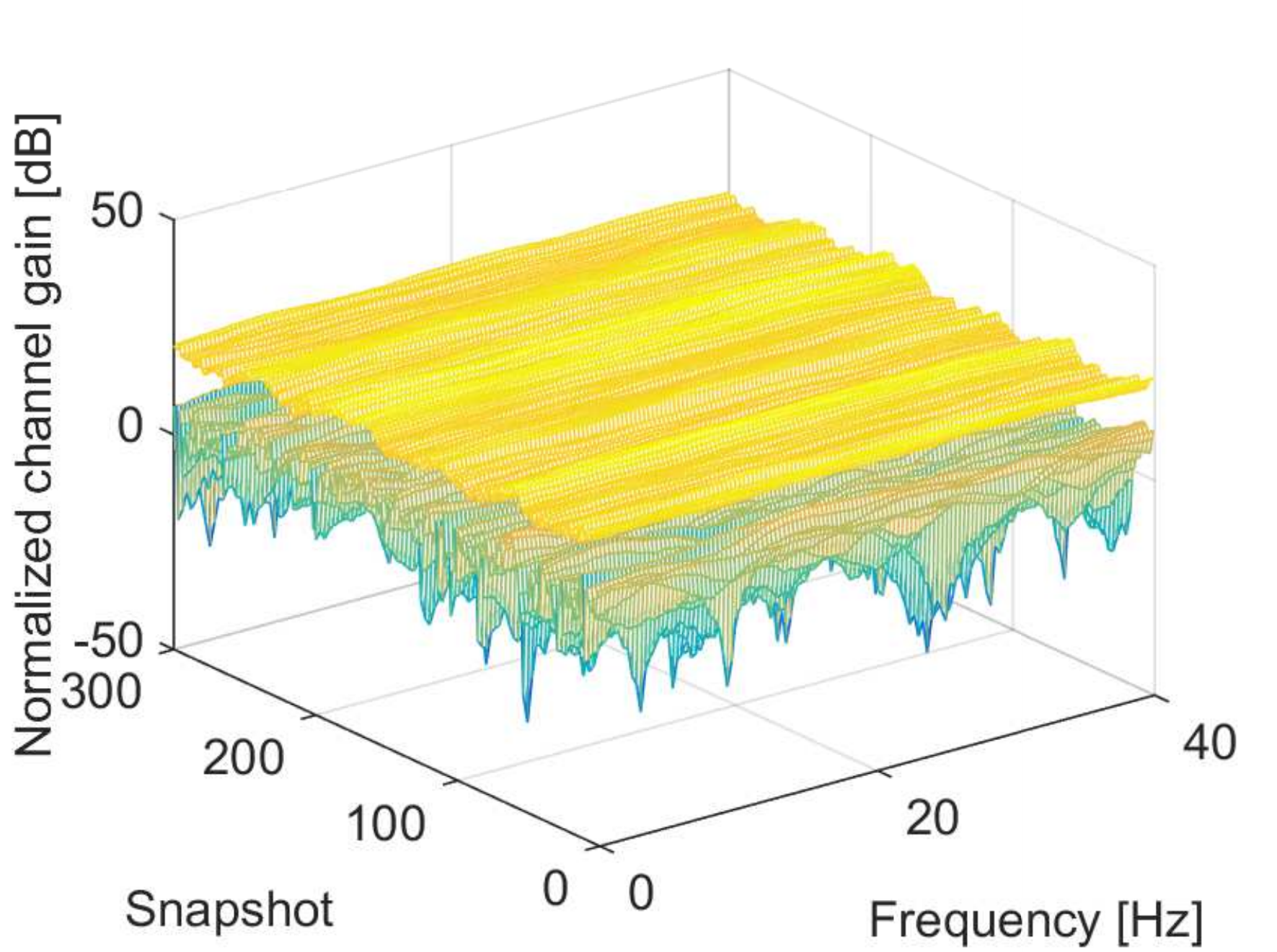}
	\caption{Normalized channel gain for one (lower) or 128~base station antennas (upper), respectively. The single base station antenna used is the one with the highest mean channel gain for user~5.}
	\label{fig:3D_best_5}
\end{figure}

\begin{figure}[t]
	\centering
	\includegraphics[width=3in]{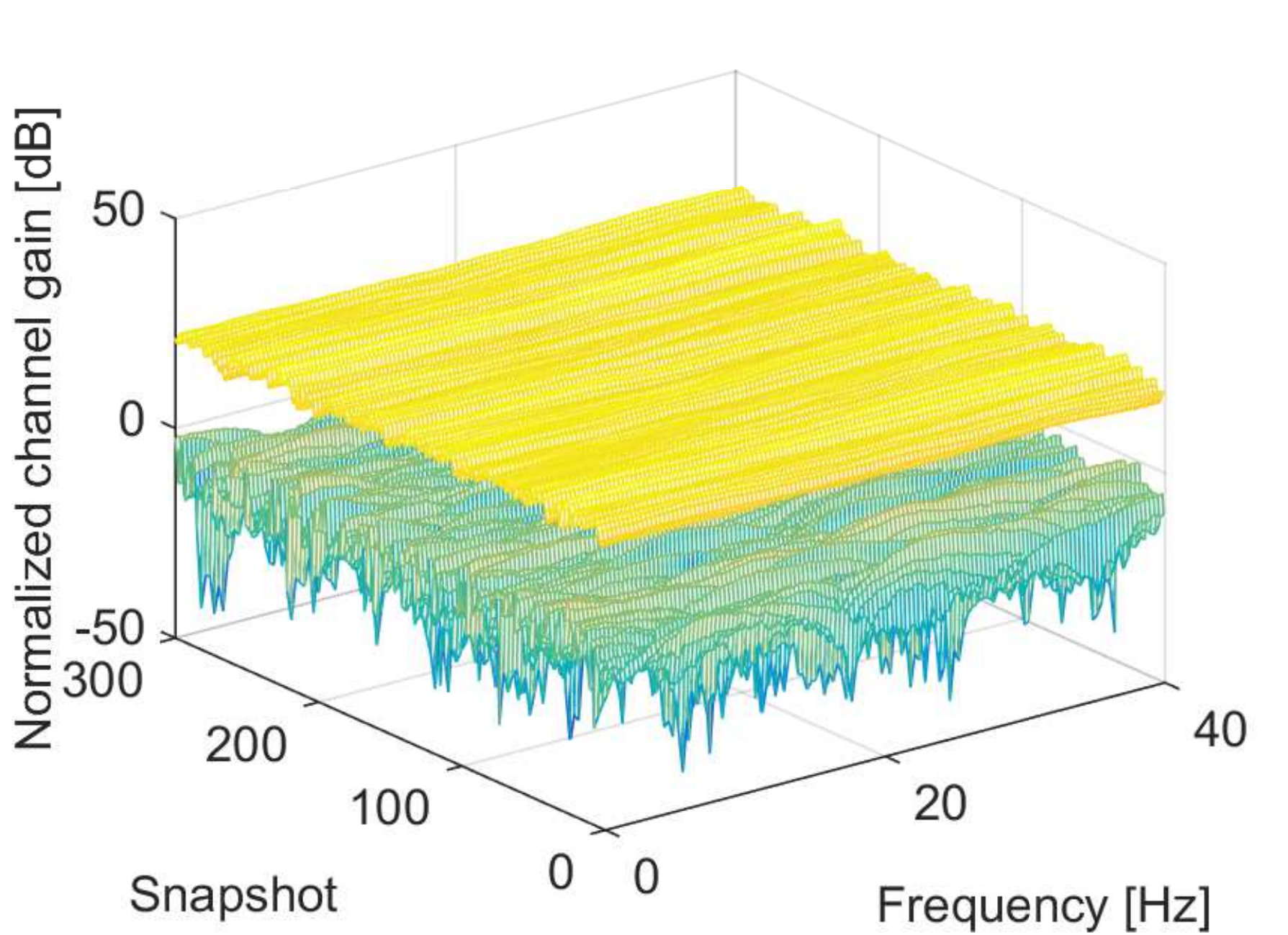}
	\caption{Normalized channel gain for one (lower) or 128~base station antennas (upper), respectively. The single antenna is the one with the lowest mean channel gain for user~1.}
	\label{fig:3D_worst_1}
\end{figure}

\begin{figure}[t]
	\centering
	\includegraphics[width=3in]{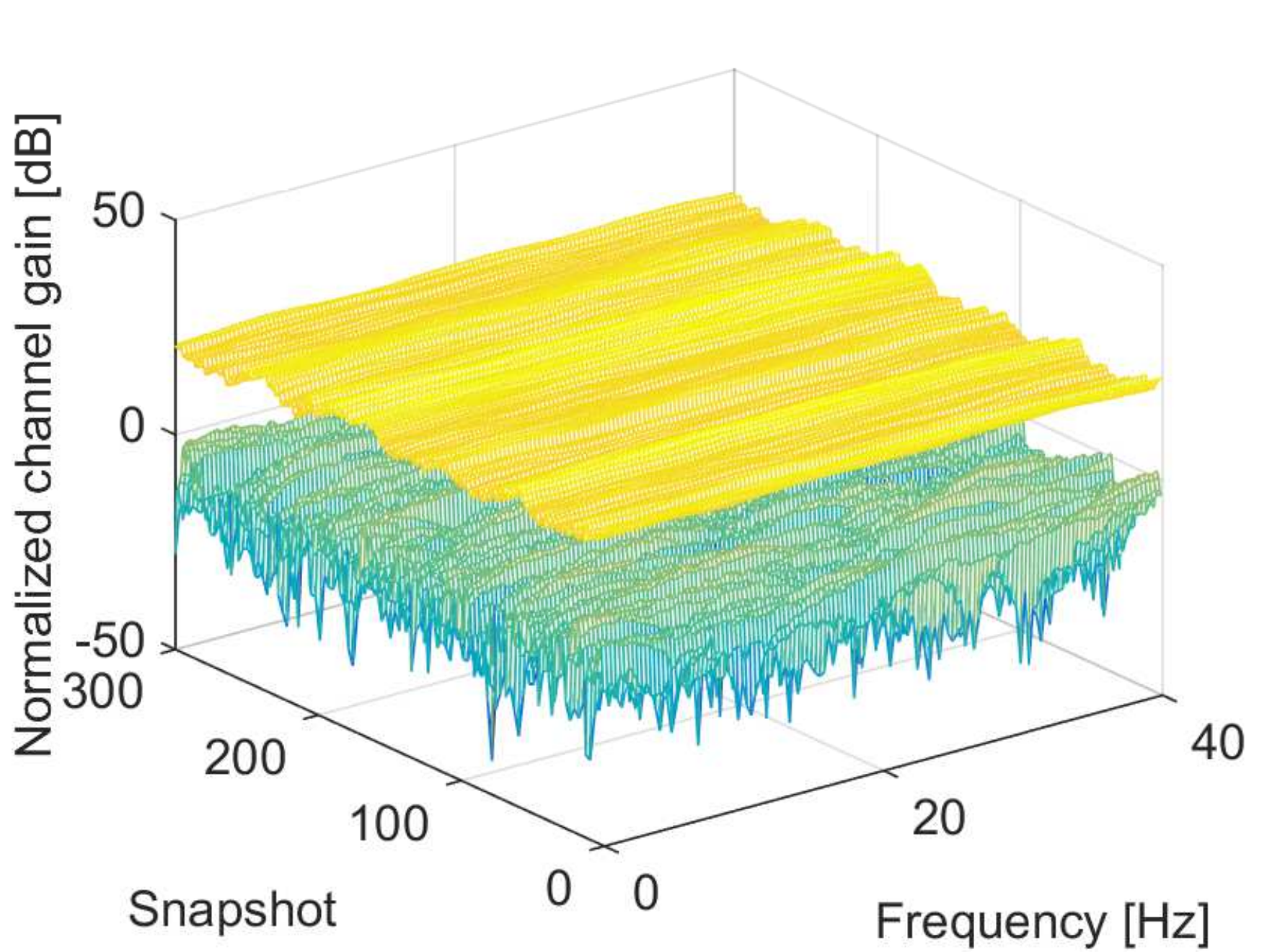}
	\caption{Normalized channel gain for one (lower) or 128~base station antennas (upper), respectively. The single antenna is the one with the lowest mean channel gain for user~5.}
	\label{fig:3D_worst_5}
\end{figure}

Continuing the evaluation of user~1 and user~5, the normalized channel gains when using one antenna versus all 128~base station antennas are shown in Figs.~\ref{fig:3D_best_1}-\ref{fig:3D_worst_5}. The normalization is performed according to (\ref{eq:norm}) with $M=128$. This is shown for the base station antenna with the highest channel gain for user~1 in Fig.~\ref{fig:3D_best_1} and for user~5 in Fig.~\ref{fig:3D_best_5}, and for the base station antenna with the lowest channel gain for user~1 in Fig.~\ref{fig:3D_worst_1}, and for user~5 in Fig.~\ref{fig:3D_worst_5}.
For the antennas with the highest channel gain, Figs.~\ref{fig:3D_best_1}-\ref{fig:3D_best_5}, it can be noticed that this single antenna has LOS condition and therefore rather large coherence bandwidth, nevertheless, this antenna also has fading dips. User~5 has a slightly more varying pattern in comparison to user~1, with reasons probably being a worse seating in the group and therefore experiencing more interaction from other persons.
Similarly to Figs.~\ref{fig:3D_best_1}-\ref{fig:3D_best_5}, the channel gain for the base station antenna with the lowest channel gain is shown in Figs.~\ref{fig:3D_worst_1}-\ref{fig:3D_worst_5} in comparison to the gain of the full channel vector. For both users there are many severe dips but for user~5 it looks slightly worse than for user~1.
What is common is the gain of the full channel vector, which is varying around $10\log_{10}(128)=21$~dB. Over frequency, the variations are quite small whereas the variations over snapshots are larger. These variations in time are created due to movements made by the users. Another observation is that the gain of the channel vector for user~5 is varying more than for user~1.

\begin{figure}[t]
	\centering
	\includegraphics[width=3in]{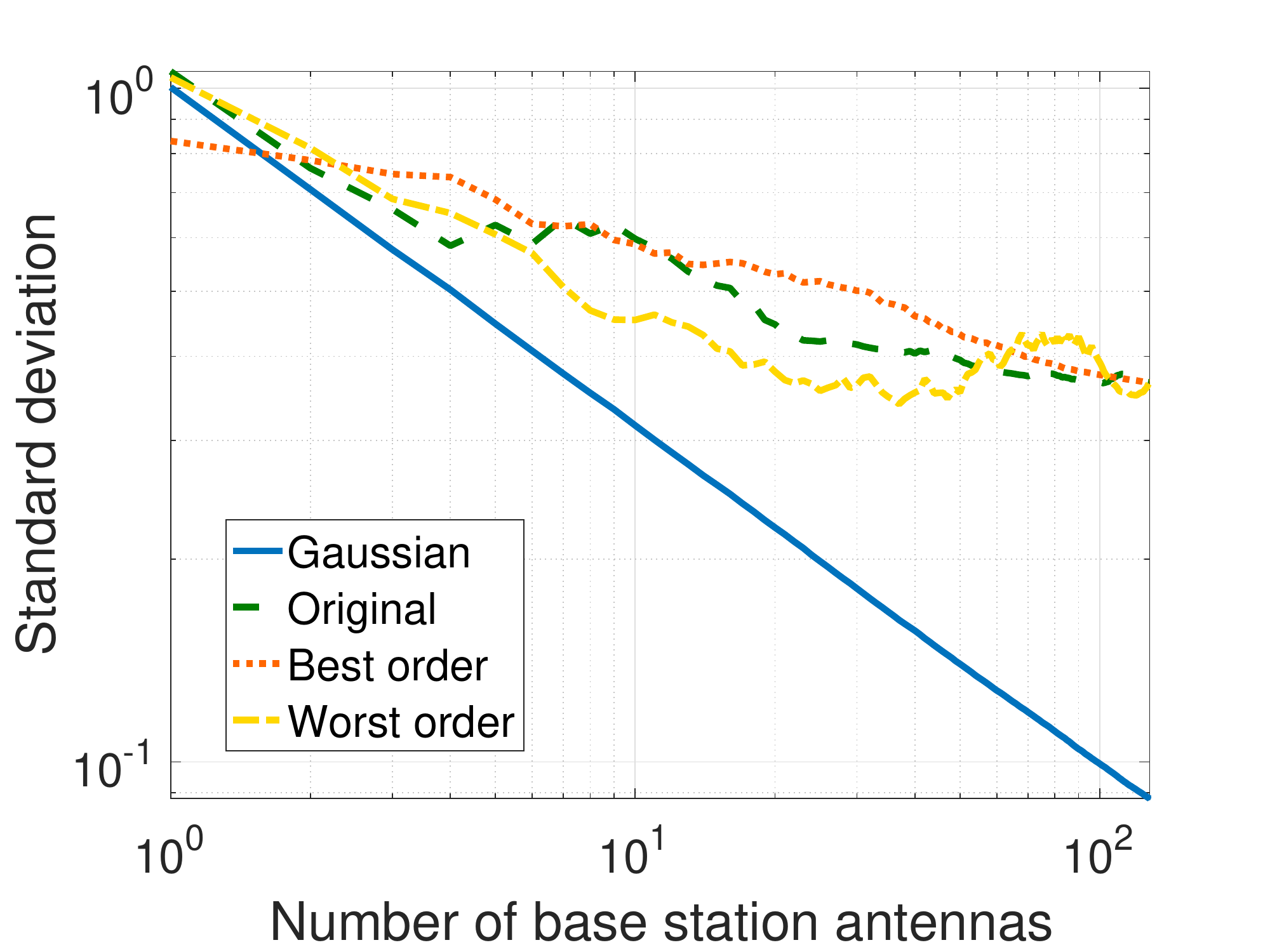}
	\caption{Standard deviation of channel gain as a function of the number of base station antennas for the Gaussian channel and user~1, when choosing the antennas in different orders.}
	\label{fig:comparison1}
\end{figure}

\begin{figure}[t]
	\centering
	\includegraphics[width=3in]{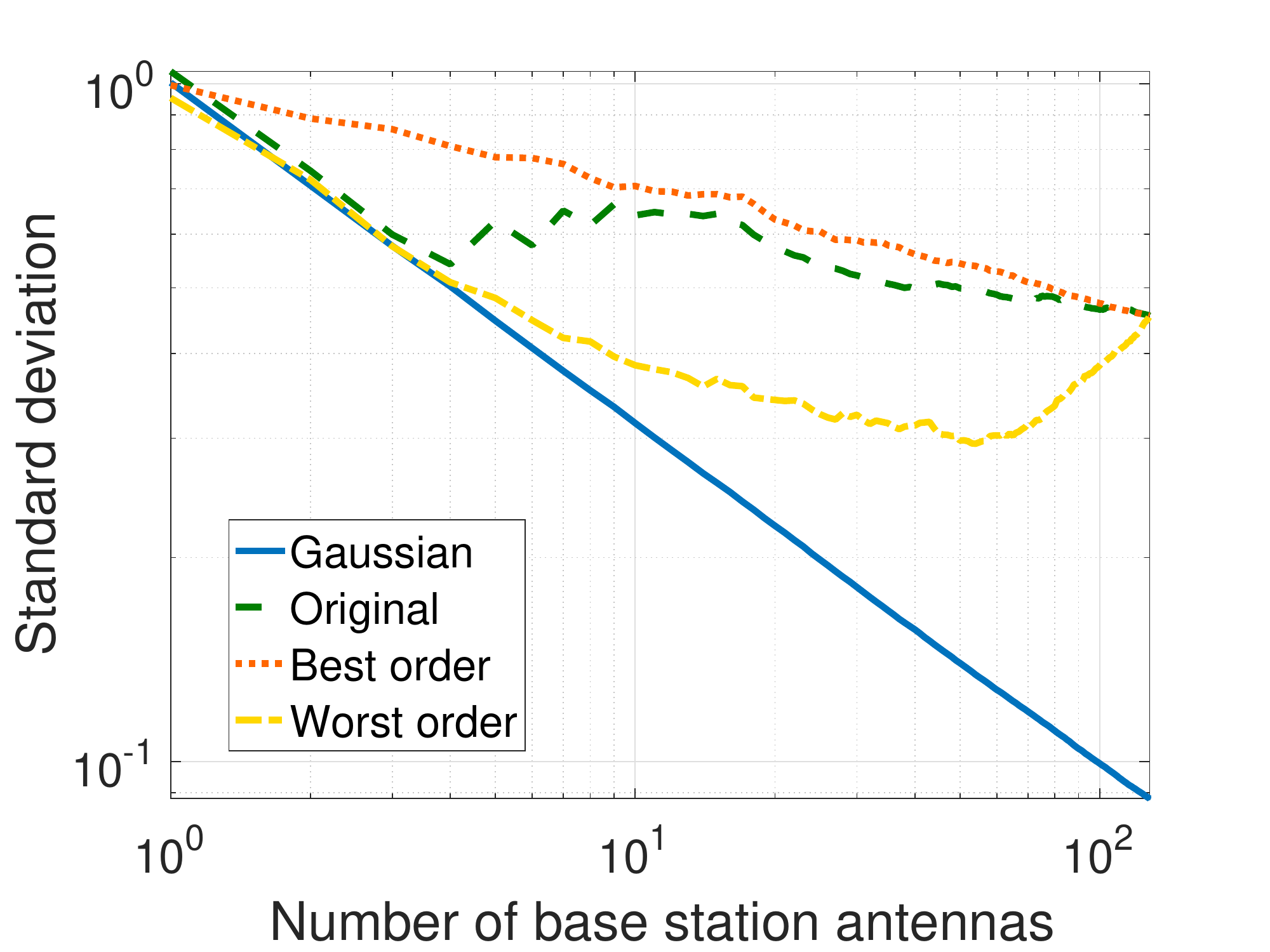}
	\caption{Standard deviation of channel gain as a function of the number of base station antennas for the Gaussian channel and user~5, when choosing the antennas in different orders.}
	\label{fig:comparison5}
\end{figure}

Further on, the standard deviations of channel gain as a function of the number of base station antennas are computed. Fig.~\ref{fig:comparison1} shows the standard deviations for user~1 and  Fig.~\ref{fig:comparison5} shows the standard deviation for user~5, both when selecting the base station antennas in different orders. For the various subsets of antennas of size $M=1,\ldots,128$, the channels are normalized according to (\ref{eq:norm}). Then, for each user, the instantaneous channel gain for every subset is computed for all frequencies and snapshots as in (\ref{eq:subset_pwr}). The standard deviation of the channel gain for each subset is computed according to (\ref{eq:std}).
The resulting standard deviation for the complex independent identically distributed (i.i.d.) Gaussian channel is plotted in both figures, with a blue solid line. The difference of the standard deviation when using 128~antennas and 1~antenna is just over $10.5$~dB, close to its theoretical value of $10\log_{10}(\sqrt{128})$. The other three curves demonstrate the measurements when choosing the antenna elements in different orders. Worth noting is that in both Fig.~\ref{fig:comparison1} and Fig.~\ref{fig:comparison5}, the mean and standard deviation are computed over all snapshots, meaning that the plots do not only show the standard deviation due to small-scale fading but also the large-scale fading caused by the interaction with the users and different antenna alignments. 

The label 'original' means that the antennas are chosen in the order as they are in the data set, described in Fig.~\ref{fig:array}. This means that the first few antennas chosen have NLOS and after that, the next few antennas have LOS and then this alternates when traversing the different rings in the array. An effect of this can be seen in both figures as the slope of the green dashed line goes steadily down in the beginning before some stronger components become a part of the subset and increase the standard deviation of the chosen subset.

The 'best order' means choosing the antennas starting with the antenna with the highest mean channel gain and the last antenna added to the subset, which includes all 128~antennas, is the antenna with the lowest mean channel gain. This means that the subsets in the beginning only includes LOS antennas and NLOS antennas are included later on. When choosing the antennas in this order, the curve goes downwards all the way. 

One thing that can be noted in Fig.~\ref{fig:comparison1} is that the standard deviation for the best antenna starts below one when having a single antenna. This is likely because this antenna experiences mostly LOS and therefore the channel gain shows less fading over frequency and time in comparison to the i.i.d. Gaussian channel, compare to Fig.~\ref{fig:3D_best_1}. For user~5, Fig.~\ref{fig:comparison5}, there are more variations in the channel, probably due to large-scale fading caused by other users since this user is placed on the back row. This can also be seen by comparing Figs.~\ref{fig:3D_best_1} and \ref{fig:3D_best_5}, where there are larger variations over time for user~5 than for user~1. 

Lastly, the 'worst order' is simply the option of choosing the antennas in the reverse order of the 'best order', meaning that the NLOS antennas are the antennas chosen first and the LOS antennas come afterwards. The result of choosing the antennas in this order is that first the standard deviation steadily decreases until the stronger antennas are included in the subset, then the standard deviation increases again, ending up at the same point as the previous antenna orders. 

Channel hardening in the measured channels depends on which order the antennas are chosen. The curves end up at the same end point but the starting point is different depending on the behavior over time for the first chosen antenna. As an attempt to quantify the channel hardening, i.e. the difference in standard deviation between, e.g., having 128~antennas and 1~antenna, varies between $3.6$~dB and $4.6$~dB for user~1, the lower one being the case when choosing the 'best order'. For user~5 the channel hardening varies between $3.2$~dB and $3.6$~dB. The difference between the two ending points for user~1 and user~5 is around $0.9$~dB. Another observation from Figs.~\ref{fig:comparison1} and \ref{fig:comparison5} is that when choosing the antennas in the 'worst order', the channel might even soften as opposed to harden, with the given normalization.

The analysis presented is based on one specific scenario, for other scenarios similar results are expected but will depend on parameters such as distribution of clusters. For future work there are several parameters that can be further examined in order to really characterize channel hardening in practice. These parameters include a further analysis of LOS/NLOS and the Ricean K-factor, polarization, different array structures and distributed arrays.

\section{Conclusion}
This paper presents a measurement-based evaluation of channel hardening in a practical scenario. The measurements were taken in an indoor auditorium with a cylindrical array, implying that some antennas are in LOS and some NLOS. The amount of channel hardening that can be expected when increasing the number of base station antennas is in this scenario highly dependent on the order in which the antennas are chosen. Depending on whether the antennas in the chosen subset are in LOS or in NLOS, both the starting point for a single antenna as well as the slope of the standard deviation curve are affected due to the large variations of channel gain over the cylindrical base station array. Also, even if the number of antenna elements at the base station side is $128$, the number of actually effective channels is less than that. Another important point in this analysis is that here, the standard deviation measured is still a result of both small-scale and large-scale fading due to interaction with the users and antenna alignments. This affects both the starting point and the slope of the standard deviation curve. Overall, based on the analysis and the specific scenario in this paper, the channel hardening, in terms of decrease of the standard deviation of the experienced channel gain, varies between $3.2$--$4.6$~dB, depending on the user's position and the order in which the antenna elements are chosen. This can be compared to the Gaussian case, where a channel hardening of $10.5$~dB is expected. Future work will include extending this analysis, to further narrow down the parameters which creates channel hardening in a practical scenario.

\section*{Acknowledgment}
This work has received funding from the strategic research area ELLIIT.

\bibliographystyle{IEEEtran}

\bibliography{my_bib}

\end{document}